\documentclass[journal=jpclcd,manuscript=letter]{achemso}

\usepackage{chemformula} % Formula subscripts using \ch{}
\usepackage[T1]{fontenc} % Use modern font encodings
\usepackage{float}
\usepackage{dcolumn}
\usepackage{graphicx}
\usepackage{siunitx}
\usepackage{amsmath}
\usepackage{amssymb}
\usepackage{braket}
\usepackage{float}
\usepackage{graphicx}
\usepackage{siunitx}
\usepackage{amsfonts}
\usepackage{color}
\usepackage{soul}
\usepackage{braket}
\usepackage{verbatim}

\newcommand{\beq}{\begin{equation}}
\newcommand{\eeq}{\end{equation}}

\author{Giovanni Batignani}
\affiliation[Rome]
{Dipartimento di Fisica,~Universit\'a~di~Roma~``La Sapienza",  ~Roma, ~I-00185, ~Italy}
\author{Carino Ferrante}
\affiliation[Rome]
{Dipartimento di Fisica,~Universit\'a~di~Roma~``La Sapienza",  ~Roma, ~I-00185, ~Italy}
\alsoaffiliation[IIT]
{~Istituto~Italiano~di~Tecnologia, Center for Life Nano Science @Sapienza, Roma, ~I-00161,  ~Italy}
\author{Giuseppe Fumero}
\affiliation[Rome]
{Dipartimento di Fisica,~Universit\'a~di~Roma~``La Sapienza",  ~Roma, ~I-00185, ~Italy}
\author{Tullio Scopigno}
\affiliation[Rome]
{Dipartimento di Fisica,~Universit\'a~di~Roma~``La Sapienza",  ~Roma, ~I-00185, ~Italy}
\alsoaffiliation[IIT]
{~Istituto~Italiano~di~Tecnologia, Center for Life Nano Science @Sapienza, Roma, ~I-00161,  ~Italy}
\email{tullio.scopigno@roma1.infn.it}

\title{Broadband Impulsive Stimulated Raman Scattering based on a Chirped Detection}

\abbreviations{ISRS}

\begin{document}

\begin{tocentry}

\includegraphics[width=1\textwidth]{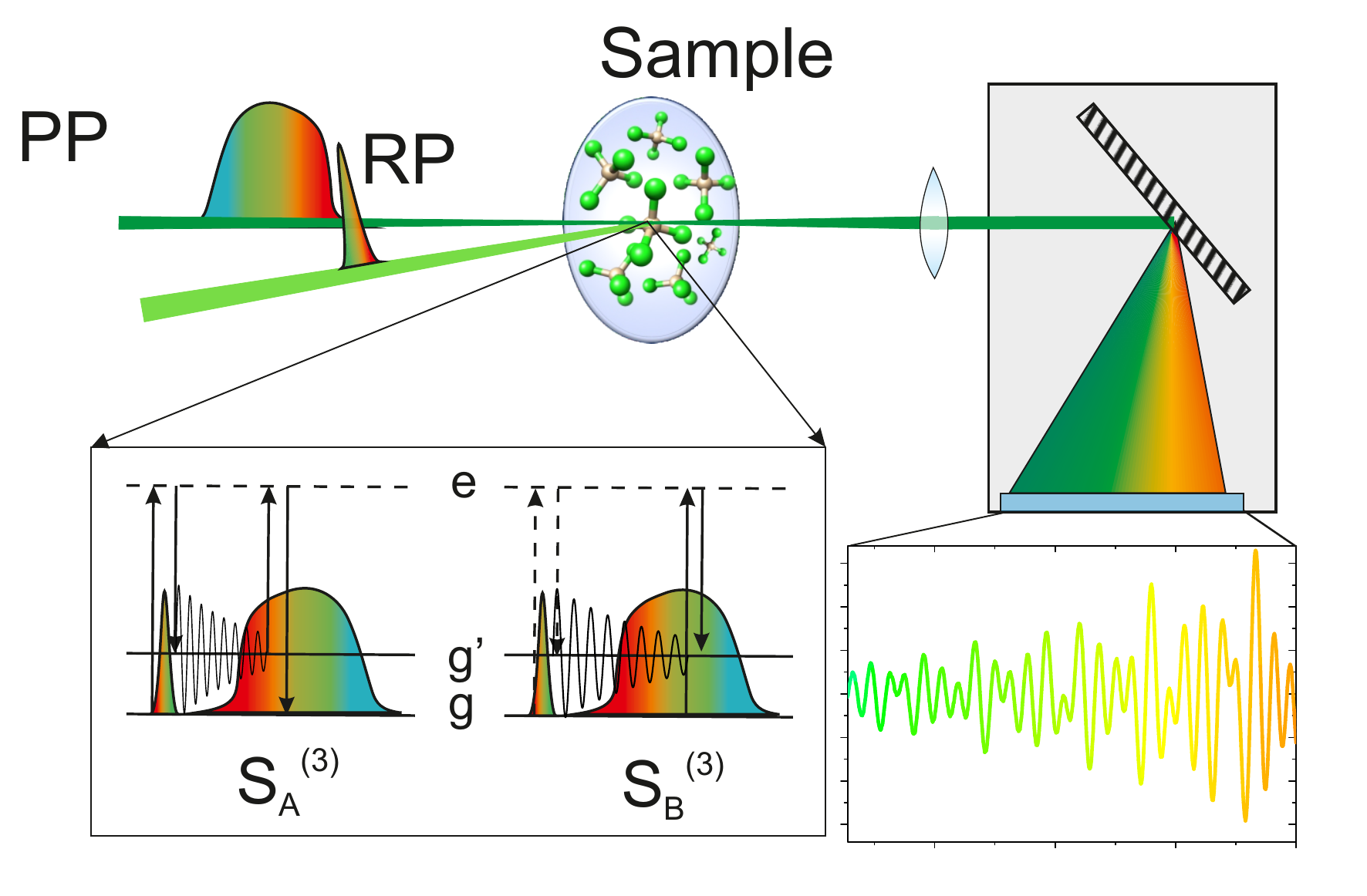}

\end{tocentry}

\begin{abstract}
	In Impulsive Stimulated Raman Scattering vibrational oscillations, coherently stimulated by a femtosecond Raman pulse, are real time monitored and read out as intensity modulations in the transmission of a temporally delayed probe pulse.
	Critically, in order to retrieve broadband Raman spectra, a fine sampling of the time delays between the Raman and probe pulses is required, making conventional ISRS ineffective for probing irreversible phenomena and/or weak scatterers typically demanding long acquisition times, with signal to noise ratios that crucially depend on the pulse fluences and overlap stabilities.
	To overcome such limitations, here we introduce Chirped based Impulsive Stimulated Raman Scattering (CISRS) technique. Specifically, we show how introducing a chirp in the probe pulse can be exploited for recording the Raman information without scanning the Raman-probe pulse delay. 
	Then we experimentally demonstrate with a few examples how to use the introduced scheme to measure Raman spectra.
\end{abstract}


Impulsive Stimulated Raman Scattering (ISRS) is a powerful technique able to monitor in time-domain vibrational fingerprints of solid state or molecular compounds using femtosecond broadband pulses. 
Within the ISRS experimental scheme, two temporally separated laser fields, conventionally referred to as  Raman pulse (RP) and probe pulse (PP), are exploited to stimulate and read out vibrational signatures in the system of interest~\cite{cit::IVS::kukura}. 
When the RP is shorter than the period of a normal mode, it can generate a localized wave-packet that coherently oscillates and evolves as a function of time. 
The photo-excited wave-packet modulates the transmissivity of the sample at the frequencies of the stimulated Raman modes, which can hence be detected by monitoring the PP transmission as a function of both temporal delay $T$ between the pulses and the probe wavelength $\lambda$. 
After Fourier transformation over $T$, ISRS yields the Raman spectrum of the system of interest.  

\begin{figure*}
	\centerline{\includegraphics[width=16cm]{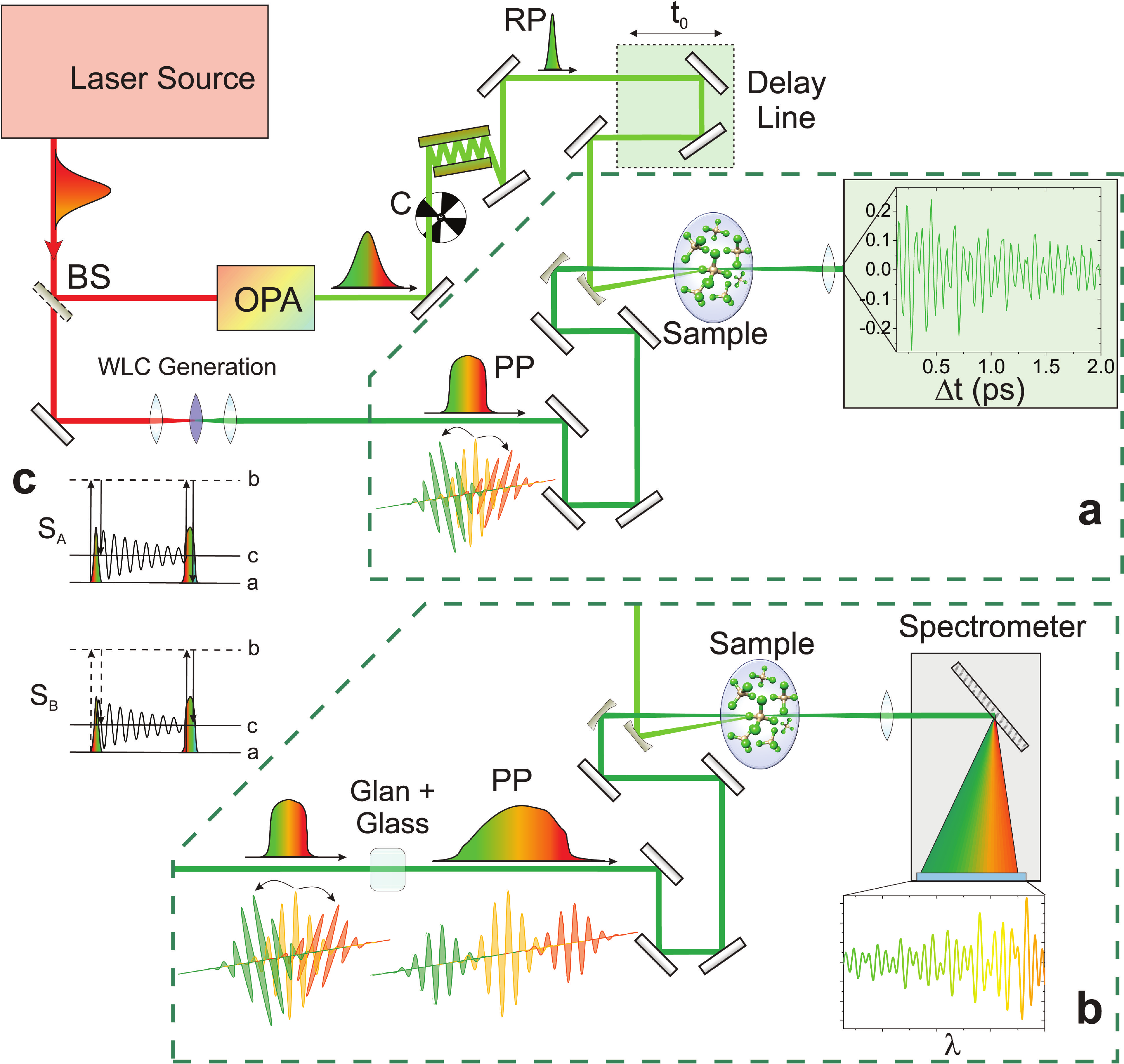}}
	\caption{Schematic layout with a comparison between conventional (a) and chirped-based (b) ISRS experimental setups. Conventional ISRS requires to measure the modified sample transmission for several time delays between RP and PP and an optical delay line is used for this purpose, while CISRS exploits a chirp in the PP to encode in the PP spectrum the same time-resolved  information on the Raman vibrations.  OPA (Optical Parametric Amplifier), RP (Raman Pump), PP (Probe Pulse), BS (Beam Splitter), C (Chopper). Energy ladder diagrams involved in the ISRS signal generation are reported in panel (c): continuous lines represent interactions of the electromagnetic field with the ket side of the density matrix, while dashed lines with the bra side.
		\label{Setup}}
\end{figure*}

A sketch of the ISRS experimental layout is presented in Fig. \ref{Setup}a.
The heterodyne detected ISRS spectra are engraved onto the highly directional PP field, efficiently suppressing both elastic and fluorescence backgrounds. For this reason, ISRS is particularly effective for probing low frequency Raman modes. Moreover, at odds with nonlinear Raman approaches in the frequency domain, such as coherent anti-Stokes Raman scattering or femtosecond stimulated Raman scattering, vibrational information retrieved in ISRS is not hampered by background signals generated by the temporal overlap of multiple pulses~\cite{Cheng2004,Batignani2019}. 
Adding an ultrashort actinic pump or tuning the Raman pulse in resonance with an electronic transition enables to trigger photochemical processes, which can then be probed with femtosecond time precision and with the structural sensitivity peculiar of Raman spectroscopy~\cite{Hamaguchi1994,Zewail_nobel}.

During the last decade, the great potential of ISRS has been exploited for studying a broad range of phenomena, including photo-isomerization processes in retinal~\cite{cit::rhuman::bacteriorhodopsin,Schnedermann2016}, intermolecular vibrational motions in liquids~\cite{cit::rhuman::liquid},
excited-state proton transfer in fluorescent protein~\cite{Fujisawa2016},  nuclear motions in photoreceptor proteins~\cite{Liebel2013,Kuramochi2017, Maiuri_2018}, singlet exciton fission in organic semiconductors~\cite{Musser2015} and polaron formation in hybrid organic-inorganic perovskites~\cite{Ghosh2017,Batignani2018,Park2018}. 
Moreover, two-dimensional implementations of ISRS have been exploited for determining nonlinear corrections to the molecular polarizability, intramolecular vibrational anharmonicities, product-reactant correlations and solvation dynamics~\cite{Tanimura1993,Tokmakoff1997,Moran2016_perspective,Kuramochieaau4490}.

Critically, at odds with frequency-domain Raman approaches, where vibrational coherences are automatically sampled over their entire dephasing time, conventional time-domain Raman protocols such as ISRS can monitor
vibrational coherences only over a finite temporal window, with a maximum time delay ($T_M$) between RP and PP.  For this reason, the ISRS lineshapes are generally affected by a limited spectral resolution $\Delta \tilde{\nu} = \frac{1}{c\cdot T_M}$.
An additional drawback of ISRS is represented by the fact that the maximum resolvable (Nyquist) frequency is limited at half of the sampling frequency ($\nu_{Sam}=\frac{1}{\Delta T}$): therefore accessing high frequency Raman modes typically requires long acquisition times~\cite{cit::IVS::kukura}. In this respect, laser fluctuations during a $T$-scan imply signal-to-noise ratios generally lower than those obtained with frequency-domain Raman spectroscopies and for low cross section or low concentration compounds they can also  overwhelm the ISRS oscillations. 
Finally, the multi-shot acquisition procedure required in ISRS cannot be applied for the study of irreversible processes, such as phase transition or non-reversible chemical reactions, or phenomena accompanied by sample damaging. 
%%% XXXX
%qua potremmo citare exp di simon wall con detection risolta spettralmente, assorbimento transiente fatto con chirp, oscillazioni in THz regimes + CARS alla Paola Borri (?)

Optical pulses are often characterized by a chirp, i.e. a time dependence (in the carrier reference frame) of their instantaneous frequency, which is in general an unfavorable effect. It hampers, for instance, reaching the minimal time bandwidth product given by the Fourier transform limit, a critical limitation in signal transmission, time resolved spectroscopies, efficiency of non-linear processes  etc. In some cases, however, it can be exploited as a useful resource.
Chirp has been employed for performing high spectral resolution spectroscopy with broadbandwidth laser pulses by spectral focusing~\cite{Hellerer2004,RochaMendoza2008,RochaMendoza2009},
to use an optical fiber as a time domain spectrometer~\cite{Saltarelli2018}, 
for enhancing vibrational signals pertaining to different electronic states~\cite{Bardeen1995,cit::RuhmanWand,Monacelli2017,Gdor2017}, up to the recent Nobel award for the amplification of optical pulses~\cite{Strickland1985}.

Here we build on a time-to-frequency encoding scheme based on chirp~\cite{Teo2015,Waldecker2015} for measuring Raman spectra, introducing the CISRS (Chirp-based Impulsive Stimulated Raman Scattering) technique. The basic idea is to exploit a chirped probe pulse, which arrives at the sample after the RP photo-excitation, so that different probe wavelengths $\lambda$ interact with the system of interest at different time delays, $t'(\lambda)$. The evolution of the stimulated vibrational coherences, encoded in the PP spectrum, is hence acquired frequency-dispersing the probe beam by a spectrometer onto a CCD device (a sketch of the experimental layout is reported in Fig. \ref{Setup}b and compared with the conventional ISRS scheme). 
Notably, the use of a frequency dispersed detection allows accessing the Raman modes which have an oscillation period much faster than the PP time duration and does not compromise the temporal resolution of the ISRS experiment~\cite{Polli2010}. 

The electric field of a linearly chirped PP with a carrier frequency $\omega_P$ can be expressed as  
$
E_P(\omega)=E_P^0(\omega) e^{i(\omega-\omega_P)^2 C}
$, where $E_P^0(\omega)$ is the square root of the PP spectrum. We define the zero of the reference temporal axis as the arrival time of the PP carrier frequency ($t(\omega_P) = 0$). Considering a RP centered at $t_0$ - with a corresponding electric field $E_R(\omega)=E_R^0(\omega)e^{+i\omega t_0}$- the relative time delay between the RP and the different probe spectral components scales as $t'(\omega)=t_0+2 C(\omega-\omega_P)$.  
%%% mettere la probe carrier fgrequency centrata a t=0 e spostare sempre l'attinica...
By synthesizing the PP through super-continuum generation\cite{cit::Agrawal}, chirped white light continuum (WLC) pulses with a large spectral bandwidth are available. The chirp value $C$ can be  further increased by introducing a dispersive material along the beam path and tailored in order to encode in the PP spectrum vibrational information sampled over a wider temporal windows. Another convenient way to add a large amount of chirp in the PP is to place along the PP optical path a Glan-laser polarizer, which can be also exploited for controlling the relative polarization between the Raman and probe pulses. For this reason, we used a 1 cm thick Glan-laser calcite polarizer.

\begin{figure*}
	\centerline{\includegraphics[width=15cm]{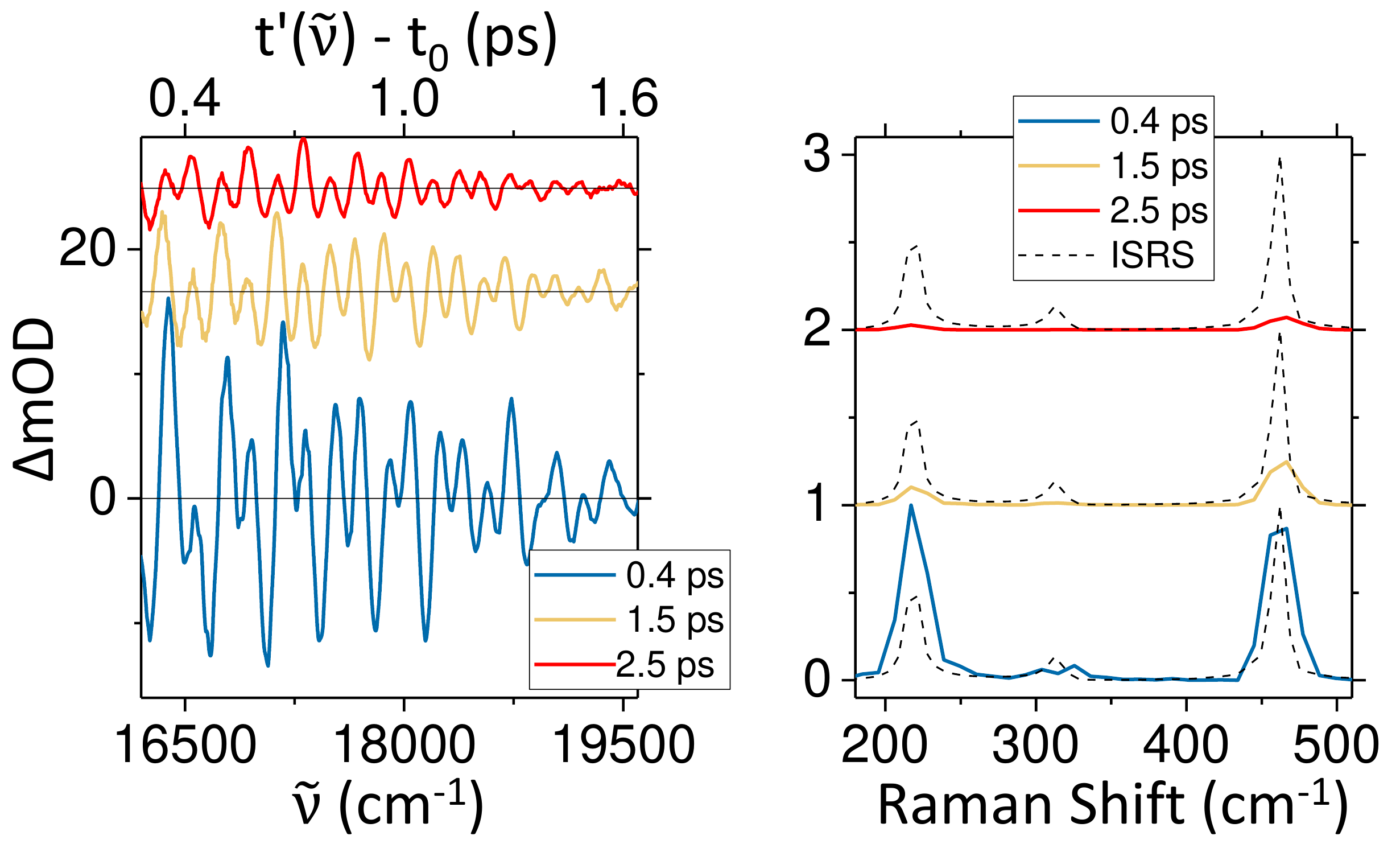}}
	\caption{Carbon tetrachloride CISRS experimental signal as a function of the probe wavenumber $\tilde{\nu}$, with the corresponding delay $t'$ between the different spectral components reported in the top axis (details on the $t'(\tilde{\nu})$ calibration are reported in the Supporting Information). Three spectra are acquired as a function of the RP temporal position $t_0$.  In the right panel we report the normalized raw Fourier transform of each $S(t',t_0)$ signal, compared with a conventional ISRS measure, normalized at the amplitude of the 460 cm$^{-1}$ peak, obtained scanning the temporal delay between Raman and probe pulses (black dashed line). Traces acquired at different $t_0$ have been vertically offsetted by a constant factor.
		\label{Fit_Wrong}}
\end{figure*}

In the left panel of Fig. \ref{Fit_Wrong} we report the CISRS signal $S(\tilde{\nu})$ acquired in a common solvent, namely liquid Carbon tetrachloride ($\mathrm{CCl_{4}}$), as a function of the probe frequency $\tilde{\nu}$ around the more stable PP spectral region (from 16500 to 19500 cm$^{-1}$). The adimensional signal $S(\tilde{\nu})$ can be defined, factoring out the probe fluence, as the difference between the transmitted probe spectrum in presence ($I_{On}(\tilde{\nu})$) and in absence ($I_{Off}(\tilde{\nu})$) of the RP, normalized by the probe pulse energy $I_{Off}$
\begin{equation}\label{eq:ISRS_Signal}
S(\tilde{\nu})=\frac{\left[I_{On}(\tilde{\nu})-I_{Off}(\tilde{\nu})\right]}{I_{Off}}
\end{equation} 
In the top axis we report the corresponding $t'(\tilde{\nu})-t_0$ axis. Considering the 16500-19500 cm$^{-1}$ PP spectral region, a $C =$ 900 fs$^2$ chirp corresponds to $\approx$ 1 ps of temporal window probed in a single-shot CISRS measurement. In order to further increase the monitored temporal region, we can tune the relative delay between Raman and probe pulses for acquiring additional CISRS spectra or increase the thickness of the dispersive material along the PP direction.
\\
In the right panel of Fig. \ref{Fit_Wrong}, the CISRS signal obtained upon Fourier transformation over $t'(\tilde{\nu})$ is reported and compared with a a conventional ISRS experiment performed on the same sample. Critically, while the position of two Raman bands (at 219 and 460 cm$^{-1}$) is in agreement with the conventional ISRS measure, the 315 cm$^{-1}$ peak is not resolved. Furthermore the lineshapes and the obtained relative intensities are not in agreement with the conventional ISRS spectrum.
\\
In order to understand the origin of such anomalous response, the expected CISRS signal can be theoretically evaluated building on a perturbative expansion of the density matrix in powers of the electric fields acting on the sample~\cite{cit::Mukamel,Mukamel_Rahav,Dorfman2013,Batignani2015pccp,Batignani2016}. In the case of non-resonant pulses, only two terms $S_A^{(3)}$ and $S_B^{(3)}$ are required to calculate the CISRS repsonse. The corresponding energy level diagrams, representing the evolution of the density matrix upon the different interaction with the RP and PP fields, are reported in Fig. \ref{Setup}c.
%While for diagram $S_A^{(3)}$ both RP and PP act on the ket side of the density matrix, the $S_B^{(3)}$ signal is generated upon  
% 
%
%
The frequency-dispersed heterodyne detected CISRS $S^{(3)} (\omega)$ signal can be expressed as the imaginary part ($\Im$) of the product between the third-order polarizations $\left[P_A^{(3)}(\omega)+P_B^{(3)}(\omega)\right]$ induced in the sample by the RP and PP fields and the complex conjugate of the probe field $E_P^*(\omega)$, accordingly to
\beq\label{Eq:Signal_Polarization_Times_Field}
S^{(3)}(\omega)=-\Im \left\{\left[P_A^{(3)}(\omega)+P_B^{(3)}(\omega)\right]E_P^*(\omega)\right\}=S_A^{(3)}(\omega)+S_B^{(3)}(\omega)
\eeq

Denoting the frequency difference between the vibrational levels $i$ and $j$ as 
$\omega_{ij}=\omega_i -\omega_j$ and the vibrational dephasing rate of the induced coherence $\ket{i}\bra{j}$ as
$\Gamma_{ij}$, with  $\tilde{\omega}_{ij}=\omega_i-\omega_j-i\Gamma_{ij}$~\cite{cit::Mukamel,Batignani2016SciRep},
$S_A^{(3)} (\omega)$ and $S_B^{(3)} (\omega)$ can be expressed as
\begin{align}\label{Eq:SA}
S^{(3)}_A (\omega) = 
\Im \Bigg[
RG_c
\int_{-\infty}^{+\infty}d\Delta
\frac{I_A (\Delta)E_P(\omega-\Delta) E_P^*(\omega)}{(\Delta-\tilde{\omega}_{ca})(\omega-\tilde{\omega}_{ba})}
\Bigg]
\end{align}
and
\begin{align}\label{Eq:SB}
S^{(3)}_B (\omega) = &\Im \Bigg[
RG_c
\int_{-\infty}^{+\infty}d\Delta
\frac{I_B (\Delta)E_P(\omega+\Delta) E_P^*(\omega)}{(-\Delta-\tilde{\omega}_{ac})(\omega-\tilde{\omega}_{bc})}
\Bigg]
\end{align}
with
\beq \label{Eq: IA,IB}
I_A (\Delta) = 
\int_{-\infty}^{+\infty}d\omega_1 
\frac{E_R(\omega_1)E_R^*(\omega_1-\Delta)}{(\omega_1-\tilde{\omega}_{ba})}
\hbox{, }
I_B (\Delta) = 
\int_{-\infty}^{+\infty}d\omega_1 
\frac{E_R^*(\omega_1)E_R(\omega_1-\Delta)}{-\omega_1-\tilde{\omega}_{ab}}
\eeq
$RG_c = \frac{\mu_{ab}\mu_{bc}\mu_{cb}\mu_{ba}}{\hbar^3}$ is a common intensity factor depending on the electronic dipole moments, with the indexes $a$, $b$ and $c$ that indicate the energy levels reported in the diagrams of Fig. \ref{Setup}c. A complete derivation of the Eqs. \ref{Eq:SA}-\ref{Eq: IA,IB} is reported in the Supporting Information (SI). 
\\
Considering a Gaussian RP $E_R^0(\omega)=e^{-(\omega-\omega_R)^2/(2s_R^2)}$ , in the electronic off-resonant regime (with $\omega_{ba}\gg \omega_R$), Eq. \ref{Eq: IA,IB} can be recast as
\begin{align*}
&I_A (\Delta) \propto 
-
\int_{-\infty}^{+\infty}d\omega 
e^{-(\omega-\omega_R)^2/(2s_R^2)+i \omega t_0} \cdot e^{-(\omega-\omega_R-\Delta)^2/(2s_R^2)-i (\omega-\Delta) t_0}=-e^{-\Delta^2/(4s_R^2)} e^{+i \Delta t_0} \\
&I_B(\Delta) \propto -e^{-\Delta^2/(4s_R^2)} e^{+i \Delta t_0}
\end{align*}
We note that the common factor $e^{-\Delta^2/(4s_R^2)}$  can be exploited to normalize the ISRS maps, in order to take into account the effect of a finite temporal duration of the RP, which impacts the ISRS response gradually reducing the Raman mode intensities upon increasing the frequency. In fact, this is done by multiplying the ISRS map for the factor $e^{\omega^2/(4s_R^2)}$. 

Defining $ss^2=\frac{s_R^2}{1-4iCs_R^2}$, the signal can be further simplified (the detailed derivation is reported in the SI) to
\begin{multline}\label{Eq:SA_Analitica}
S^{(3)}_A (\omega) \propto   
\Im \Bigg\{ 
-i\pi RG_c
E_P^{0}(\omega-\omega_{ca}) E_P^{0*}(\omega) 
e^{-(\Gamma_{ca}+i\omega_{ca})[t_0+2(\omega-\omega_P) C]} e^{\frac{(\Gamma_{ca}+i \omega_{ca})^2}{4 \text{ss}^2}}
\\  
 \left[1- \text{Erf}\left(\frac{\Gamma_{ca}-2 [t_0+2(\omega-\omega_P) C] ss^2+i \omega_{ca}}{2 ss}\right)\right]
\Bigg\}
\end{multline}
and
\begin{multline}\label{Eq:SB_Analitica}
S^{(3)}_B (\omega) \propto 
\Im \Bigg\{+i\pi RG_c E_P^{0}(\omega+\omega_{ca}) E_P^{0*}(\omega)
e^{-(\Gamma_{ca}-i\omega_{ca})[t_0+2(\omega-\omega_P) C]}
e^{\frac{(\Gamma_{ca}-i\omega_{ca})^2}{4ss^2}}
\\ 
\left[1-\text{Erf}\left(\frac{\Gamma_{ca}-2[t_0+2(\omega-\omega_P) C]ss^2-i\omega_{ca}}{2ss}\right)\right]
\Bigg\}
\end{multline}
As expected both  $S^{(3)}_A (\omega) $ and $S^{(3)}_B (\omega)$ oscillate at the $\omega_{ca}$ frequency and decay with the $\Gamma_{ca}$ rate. 
While the $S_A^{(3)} (\omega)$ signal is generated by an interaction with the PP field red-shifted by one vibrational quantum -$E_P(\omega-\omega_{ca})$- with respect to the detected probe frequency $\omega$, the  $S_B^{(3)} (\omega)$ contribution arises from the blue shifted component $E_P(\omega+\omega_{ca})$~\cite{Monacelli2017}. 
Considering a fixed $t_0$, the CISRS frequency resolved signals can be converted to the corresponding temporal delays $t'(\omega)=t_0+2\omega C$. 
Critically, Fourier transforming $S^{(3)}_A$ or $S^{(3)}_B$ over $t'$ provides a direct information on the vibrational coherence only if the PP is constant over the monitored spectral region  ($E_P^{0}(\omega)=E_P^{0}$).   
%We note that, if we convert the  $S^{(3)}_A(\omega)$-$S^{(3)}_B(\omega)$ signals detected at each probe frequency $\omega$ to the corresponding delay with respect to the RP photo-excitation $t'=t_0+2\omega C$, a Fourier transform provide a direct information 
%We note that, in order to have $S^{(3)}_A$-$S^{(3)}_B$ signals that depend only on the parameter $t_0+2\omega C$  and hence converting the measured signal at frequency $\omega$ to the corresponding temporal delay $t'=t_0+2\omega C$,  
%We note that, for having $S^{(3)}_A$ and $S^{(3)}_B$ signals that depend only on the parameter $t_0+2\omega C$,  the  monitored PP spectral region has to be constant ($E_P^{0}(\omega)=E_P^{0}$).
% questo è vero in generale anche per chirp di ordini successivi...
%% un requisito fondamentale per passare dal segnale misurato alla frequenza di probe omega al ritardo temporale associato alla componente spettrale in questione (e poter fare quindi la traformata di fourier) è necessario che 
\begin{figure*}
	\centerline{\includegraphics[width=16.5cm]{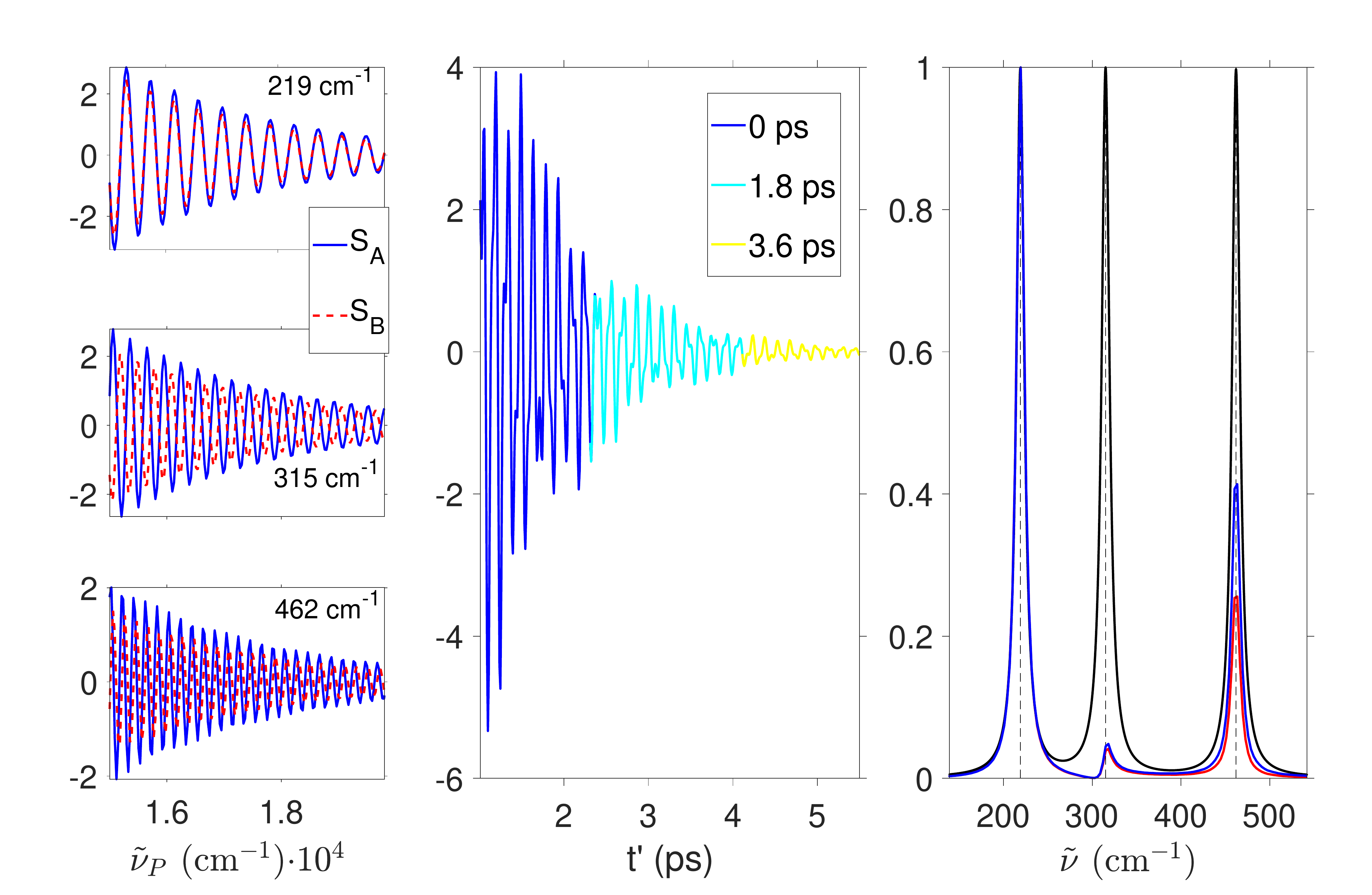}}
	\caption{$S^{(3)}_A$ and $S^{(3)}_B$ CISRS signals computed in the ideal case of a spectrally flat probe pulse for a model system with three Raman active modes at 219, 315 and 462 cm$^{-1}$, with equal amplitudes and dephasing time (1 ps), for $C$ = 945 fs$^{2}$ and a 30 fs Raman pulse. In the central panel we report the total signal as a function of $t'(\omega)=t_0+2\omega C$ for three different time delays $t_0$. In the right panel the corresponding Fourier transform (red line) is shown and is compared with the Raman spectral lineshape (black line). The blue line reports the Fourier transform spectrum corrected for the finite RP duration and has been obtained multiplying the red trace for the factor $e^{\omega^2/(4\sigma_R^2)}$.
		\label{IdealProbe}}
\end{figure*}
In Fig. \ref{IdealProbe} we show the signals $S^{(3)}_A$ and $S^{(3)}_B$ computed for such an ideal case, with a spectrally flat PP, using a 30 fs RP and a PP chirp $C$ = 945 fs$^2$.
We considered a model system with three Raman active modes at frequencies 219, 315 and 462 cm$^{-1}$, with equal amplitudes and the same dephasing time (1 ps). 
In order to increase the effective probed temporal window, and hence the spectral resolution upon Fourier transforming (right panel), we evaluated the CISRS total response (reported in the central panel and obtained as the sum of $S^{(3)}_A$ and $S^{(3)}_B$) for three  time delays $t_0$ between the Raman and the probe pulses. 
Critically,  as highlighted in the left panel, $S^{(3)}_A$ and $S^{(3)}_B$  can be in phase or out of phase depending on the chirp and on the Raman mode under investigation~\cite{Monacelli2017,Gdor2017}. Specifically, considering a mode at frequency $\omega_{ca}$, the two signals are in phase for $C=\frac{(2n+1)\pi}{2\omega_{ca}^2}$ and are out of phase for $C=\frac{n\pi}{\omega_{ca}^2}$ ($n=0,1,2,...$) for a spectrally flat PP. 
Therefore the amplitude of the ISRS oscillations can vary as a function of the probe wavelength also in the electronic off-resonant regime and in general the intensity of the ISRS peaks does not directly reflect the Raman cross-section. 
Moreover, the broadband pulses synthesized by optical parametric amplification or supercontinuum generation are in general not spectrally flat and hence a direct conversion of the CISRS spectrally detected signal to the corresponding time delay $t'(\omega)=t_0+2\omega C$  is not valid. 
\\
For these reasons, the Raman spectrum obtained upon a direct Fourier transform of the measured CISRS signal is not effective, since specific modes under investigation can be suppressed and the Raman lineshapes can be altered. 
It is worth stressing that using a non spectrally flat PP from one hand requires a more complex modeling, but from the other ensures that the destructive interference between the signals $S^{(3)}_A$ and $S^{(3)}_B$ does not cancel the vibrational response of Raman modes at $\omega_{ca}=\sqrt{\frac{n\pi}{C}}$: 
since the amplitude of the two contributions is proportional to the intensity of the probe field at different frequencies,  the $S^{(3)}_A$ and $S^{(3)}_B$ terms do not cancel each other out.

In order to correctly extract the Raman information from a spectrally dispersed CISRS trace, 
Eqs. \ref{Eq:SA}-\ref{Eq: IA,IB} should hence be used. We note that such relations are valid not only for Gaussian pulses but hold for arbitrary RP and PP spectral profiles. 
Moreover, by expanding the PP phase in power of $\omega$, accordingly to $E_P(\omega)=E_P^0(\omega)e^{i\sum_n C_n\left(\omega-\omega_P\right)^n}$, the derived formalism can be exploited for taking into account higher order dispersion terms beyond then linear chirp regime, a crucial factor for a correct calibration of the arrival time of each PP component in a broad spectral range. 
In order to estimate $C_n$ for the experimental PP, we measured the temporal overlap between the different spectral component of the broadband PP and the femtosecond RP inside the sample by means of the Optical Kerr Effect (OKE) cross-correlation~\cite{McCamant2004} (further details are reported in the Methods section and in the SI). In our case, we used a PP with $C_2$ = 837 fs$^2$ and $C_3$ = 194 fs$^3$, which correspond to a temporal window equal to $\sim 1.3$ ps in the considered spectral range.

\begin{figure*}
	\centerline{\includegraphics[width=16cm]{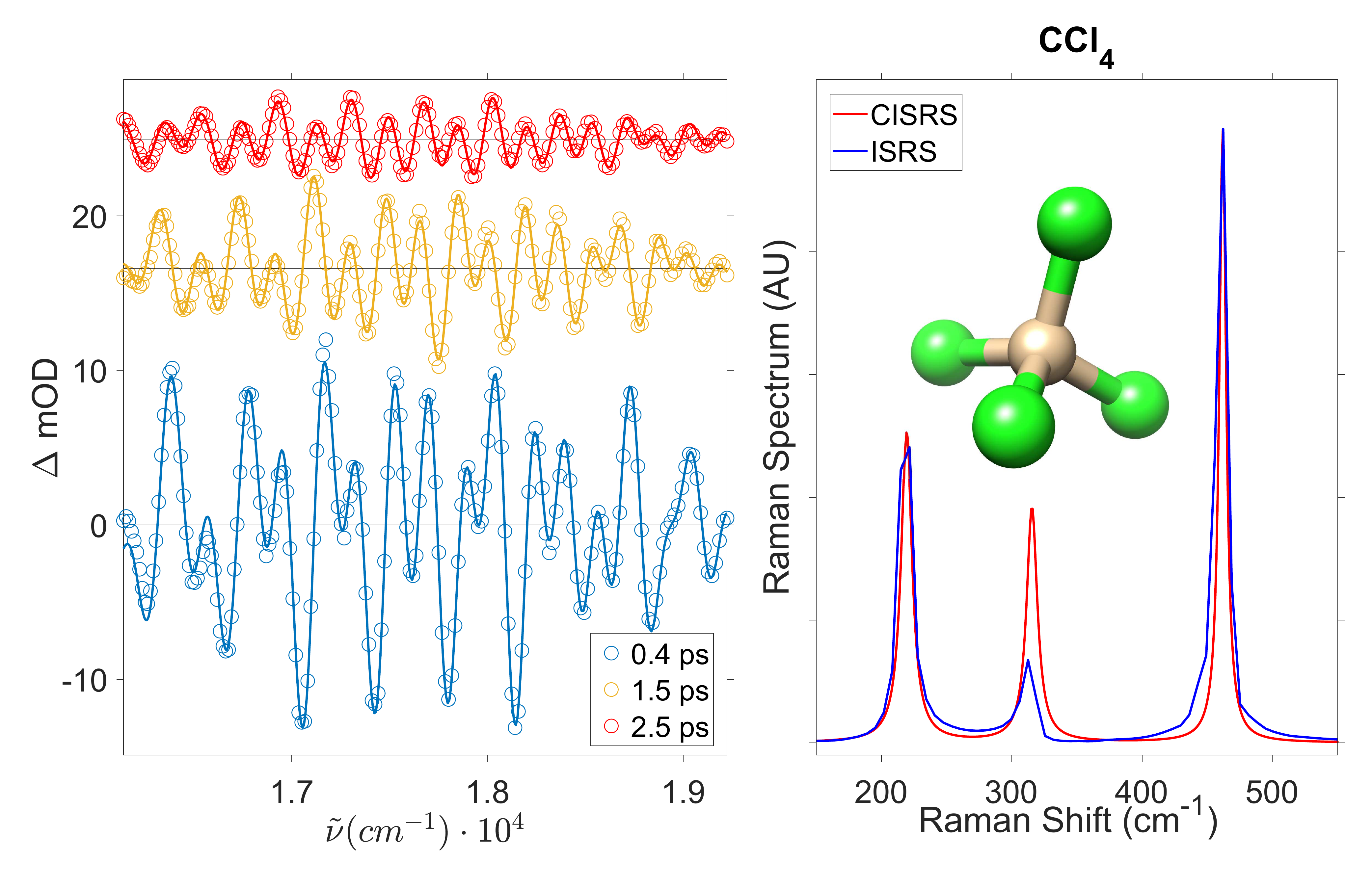}}
	\caption{CISRS experimental signal (dots) of carbon tetrachloride as a function of the probe wavenumber $\tilde{\nu}$ acquired for the three different RP delays $t_0$ reported in the box. 
	The signals modeled using  Eqs. \ref{Eq:SA}-\ref{Eq: IA,IB} are reported as continuous lines and show a good agreement with the experimental data. Different time traces have been offsetted by a constant factor.
	In the right panel we report the CISRS extracted Raman spectrum, compared with a conventional time-domain ISRS spectrum (blue line).  
		\label{Fit_CCl4}}
\end{figure*}
\begin{figure*}[h]
	\centerline{\includegraphics[width=16cm]{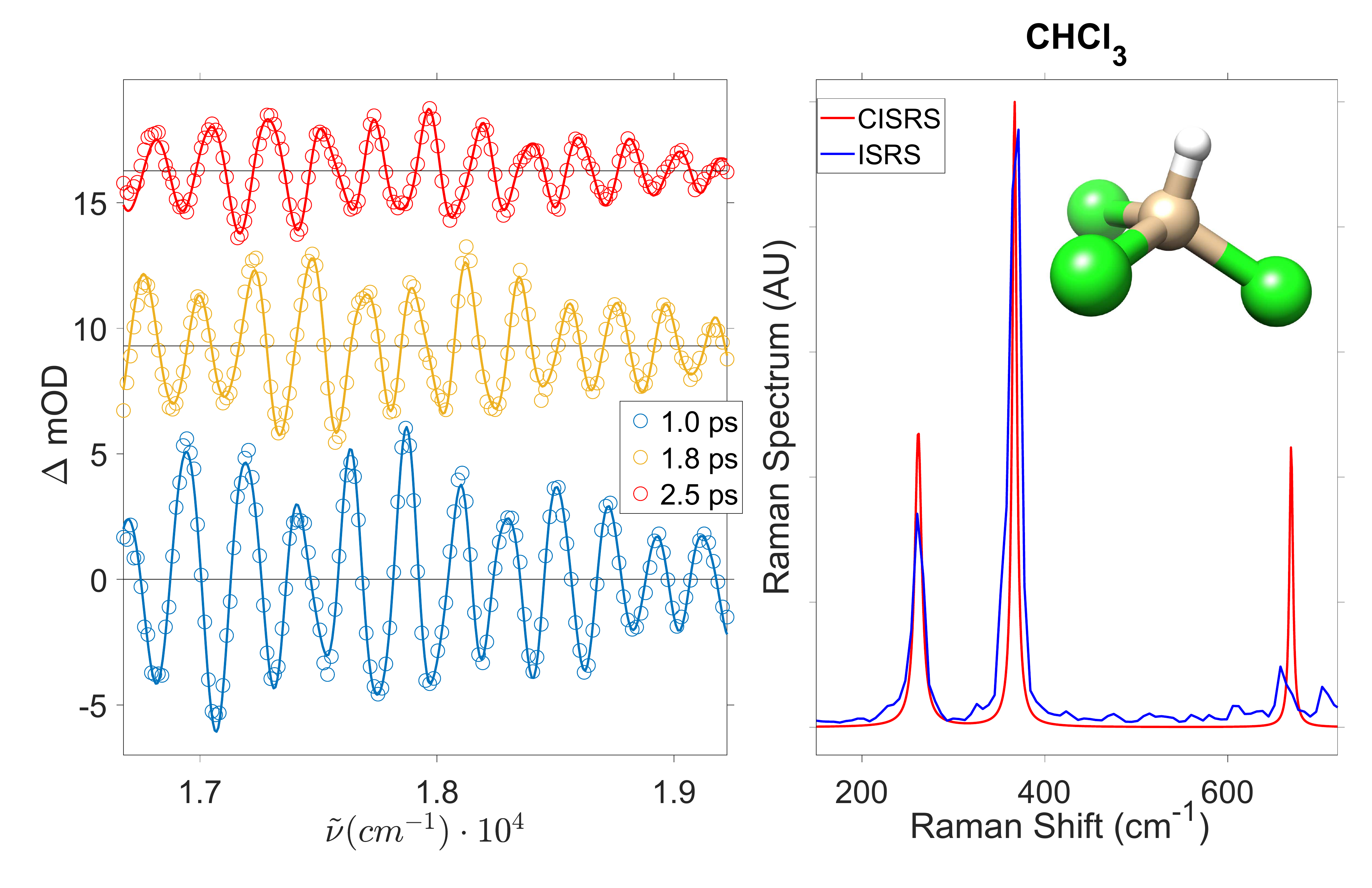}}
	\caption{CISRS experimental signal (dots) of chloroform as a function of the probe wavenumber $\tilde{\nu}$ acquired for the three different RP delays $t_0$ reported in the box. 
		The signals modeled using  Eqs. \ref{Eq:SA}-\ref{Eq: IA,IB} are reported as continuous lines and show a good agreement with the experimental data. Different time traces have been offsetted by a constant factor. In the right panel we report the CISRS extracted Raman spectrum, compared with a conventional time-domain ISRS spectrum (blue line).  
		\label{Fit_CHCl3}}
\end{figure*}

In Figs. \ref{Fit_CCl4}  and \ref{Fit_CHCl3} we report the CISRS traces of carbon tetrachloride and chloroform ($\mathrm{CHCl_{3}}$), respectively. The experimental data (dots) are in good agreement with the signal modeled using  Eqs. \ref{Eq:SA}-\ref{Eq: IA,IB} (continuous lines). 
Specifically, Eqs. \ref{Eq:SA}-\ref{Eq: IA,IB} have been exploited to fit the experimental traces in order to retrieve the matter parameters ruling the Raman scattering cross section (normal mode frequencies, dephasing times and Raman intensity), which have been reported in the right panels of Figs. \ref{Fit_CCl4}  and \ref{Fit_CHCl3}. 
The CISRS extracted spectrum is compared with the conventional ISRS signal obtained by Fourier transforming over the scanned temporal delay between the Raman and probe pulses (further details in the SI). The ISRS Raman spectrum has been corrected to take into account the finite RP time duration (30 ps).
To ensure a high spectral resolution, in the ISRS measure we used a $\sim$ 4 ps time-window, with a $\Delta T = $ 10 fs sampling step. 
\\
The CISRS measured peak positions are 219,  315.5 and  462  cm$^{-1}$ for $\mathrm{CCl_{4}}$ (with corresponding measured dephasing times equal to 1, 1.1 and 1.7 ps), and 
261.5,  367 and  669 cm$^{-1}$ for $\mathrm{CHCl_{3}}$ (1.2, 1.9 and 2.3 ps measured dephasing times). All of them are in good agreement with the frequency domain spontaneous and stimulated Raman, as well as with ISRS measurements, demonstrating the CISRS capability of  measuring Raman spectra with fast acquisition times.  
A detailed comparison between normal peak frequencies and line-widths is reported in Table \ref{table: SRS}.
%\\
Notably, while the  ISRS measurement required 400 different acquisitions for retrieving the Raman information, the CISRS was performed with only 3 steps. 
Moreover, the CISRS Raman spectrum shows an improved signal-to-noise ratio with respect to the ISRS one: in fact, the fast  CISRS acquisition time ensures that the measure is not affected by laser fluctuations or by beam dealignment, which on the contrary limit conventional ISRS.  
We note that under the present experimental conditions, a single CISRS measurement is still able to retrieve the Raman information but with a lower spectral resolution. A convenient way to increase the such resolution for performing single-shot CISRS acquisitions is introducing additional chirp or increasing the monitored PP spectral region.
\begin{comment}
\begin{center}
\begin{table}
\begin{tabular}{ |c|c||c|c| }
\hline
\multicolumn{4}{|c|}{Carbon tetrachloride}\\
\hline
\multicolumn{2}{|c||}{Peak Position (cm$^{-1}$)} & \multicolumn{2}{|c|}{FWHM (cm$^{-1}$)}\\
\hline
CISRS     & Spontaneous Raman  & CISRS     & Spontaneous Raman \\
\hline
219 & 217\cite{book_Shimanouchi}, 217\cite{Marrinan1954} & 11 & 11\cite{Gaynor2015}\\
\hline
315.5 & 315.5\cite{book_Shimanouchi}, 313\cite{Marrinan1954} & 10 & 11\cite{Gaynor2015}\\
\hline
462 & 459\cite{book_Shimanouchi}, 459\cite{Marrinan1954} & 6 & 9\cite{Gaynor2015}\\
\hline
\multicolumn{4}{|c|}{Chloroform}\\
\hline
\multicolumn{2}{|c||}{Peak Position (cm$^{-1}$)} & \multicolumn{2}{|c|}{FWHM (cm$^{-1}$)}\\
\hline
CISRS     & Spontaneous Raman & CISRS     & Spontaneous Raman \\
\hline
261.5 & 261\cite{book_Shimanouchi}, 261\cite{Marrinan1954} & 9 & 9\cite{McCreery2000}\\
\hline
367 & 363\cite{book_Shimanouchi}, 366\cite{Marrinan1954} & 6 & 5\cite{McCreery2000}\\
\hline
669 & 672\cite{book_Shimanouchi}, 667\cite{Marrinan1954} & 5 & 4\cite{McCreery2000}\\
\hline
\end{tabular}
\caption{Comparison of Raman peak positions and full width half maxima (FWHM) measured for carbon tetrachloride and chloroform by CISRS and spontaneous frequency-domain Raman spectroscopy.}
\end{table}
\end{center}
\end{comment}
%\vspace{2cm}
\begin{table}
	\begin{tabular}{ |c|c|c||c|c|c| }
		\hline
		\multicolumn{6}{|c|}{Carbon tetrachloride}\\
		\hline
		\hline
		\multicolumn{3}{|c||}{Peak Position (cm$^{-1}$)} & \multicolumn{3}{c|}{FWHM (cm$^{-1}$)}\\
		\hline
		CISRS     & Spontaneous Raman & SRS  & CISRS     & Spontaneous Raman & SRS\\
		\hline
		219 & 217\cite{book_Shimanouchi}, 217\cite{Marrinan1954} & 219&11 & 11\cite{Gaynor2015}& 10\\
		\hline
		315.5 & 315.5\cite{book_Shimanouchi}, 313\cite{Marrinan1954} & 315&10 & 11\cite{Gaynor2015} &11\\
		\hline
		462 & 459\cite{book_Shimanouchi}, 459\cite{Marrinan1954} & 460&6 & 9\cite{Gaynor2015}& 8\\
		\hline
		\hline
		\multicolumn{6}{|c|}{Chloroform}\\
		\hline
		\hline
		\multicolumn{3}{|c||}{Peak Position (cm$^{-1}$)} & \multicolumn{3}{c|}{FWHM (cm$^{-1}$)}\\
		\hline
		CISRS     & Spontaneous Raman & SRS  & CISRS     & Spontaneous Raman & SRS\\
		\hline
		261.5 & 261\cite{book_Shimanouchi}, 261\cite{Marrinan1954} & 263&9 & 9\cite{McCreery2000}& 8\\
		\hline
		367 & 363\cite{book_Shimanouchi}, 366\cite{Marrinan1954} & 367&6 & 5\cite{McCreery2000}& 4\\
		\hline
		669 & 672\cite{book_Shimanouchi}, 667\cite{Marrinan1954} & 669&5 & 4\cite{McCreery2000}& 4\\
		\hline
	\end{tabular}
	\caption{Comparison of Raman peak positions and full width half maxima (FWHM) measured for carbon tetrachloride and chloroform by CISRS and two frequency-domain Raman approaches, namely spontaneous and stimulated Raman scattering (SRS).}\label{table: SRS}
\end{table}

%Lavori Cars che usano il chirp: CPP-CARS \cite{Lang_2002}

%~\cite{Hellerer2004} %%% introdotto lo 'spectral 'focusing' nel quale si introduce lo stesso amount di chirp in pump e probe pulses, demonstrating how to achieve high spectral resolution in CARS microscopy using broad-bandwidth femtosecond pulses.

%~\cite{RochaMendoza2008,Pestov2008,RochaMendoza2009,Langbein2009} %% sostituisce grating con vetro rispetto a hellerer2004

%%% in such a way, the resulting spectral width of EPE∗S fields centered at the instantaneous frequency difference is given by the Fourier-limit of the temporal envelope of the pulses, which can be elongated by the applied linear chirp to many picoseconds. This approach additionally allows to tune the IFD within the spectral width of the pulses by simply adjusting their relative time delay, and thus to perform CARS spectroscopy without tuning the laser sources.

%%% Langbein2009: this way of introducing linear chirp is significantly simplifying the design, eliminating alignment issues, and increasing the efficiency compared to previously employed pulse shaping based on spatial light modulators and/or dispersion gratings or prisms based systems.

%%% 

In Summary, we have introduced a novel experimental scheme based on the use of a broadband chirped probe pulse to record in time-domain vibrational coherences and thereby Raman spectra.
In CISRS the signal measured at a given probe wavelength does not simply correspond to a contribution generated at the arrival time of the corresponding PP spectral component, but rather it results from the interference of signals generated at different time delays. 
Building on a perturbative expansion of the density matrix representing the matter evolution during sequential interaction with the fields, we have shown how to retrieve the Raman information encoded in the spectrally dispersed probe. 
At odds with conventional ISRS, the presented scheme does not require to scan the temporal delay between pump and probe pulses, ensuring acquisition times two orders of magnitude faster, with comparable or better temporal and spectral resolutions. 
Moreover, the chirped based scheme is not affected by pump instabilities or by beams dealignments which occurs during the time-delay scan required by conventional ISRS, ensuring improved signal-to-noise ratios.
The presented scheme can be exploited for mapping the ultrafast Raman response of irreversible processes, such as phase transition, non-reversible chemical reactions or phenomena accompanied by sample damaging, where on the contrary ISRS cannot be applied. 
Finally, we anticipate that the chirp based detection scheme
introduced here can be extended to multidimensional time domain Raman spectroscopy, representing a fast acquisition method, which simplifies the experimental design and at the same time increases the signal to noise ratios.
%chirping the probe pulse can be exploited also for building fast multidimensional ISRS spectroscopic setup.

%DA VALUTARE SE NELLE CONCLUSIONS METTERE CHE GRAZIE ALLA RAPIDIT\'A CON CUI COL PRESENTE SCHEMA SI POSSONO MISURARE SPETTRI ISRS, IL PRESENTE SCHEMA PU\'O ANCHE ESSERE IMPLEMENTATO COME TECNICA DI FAST IMAGING, BENEFICIANDO DI TUTTI I VANTAGGI DEL RAMAN COERENTE, COME LA SENSIBILIT\'A STRUTTURALE, LA 3D SPATIAL RESOLUTION COME CARS + non-invasiveness OFFERTA DAL RAMAN.

%For these measurements to be performed in a manageable amount of time, one spectral axis is typically recorded in a single laser shot. An analogous rapid-scanning capability for THz measurements will unlock the multidimensional toolkit in this frequency range.

\section{Methods}

\textbf{Experimental setup}.
The experimental setup exploited for the measurements is based on a Ti:sapphire laser source that generates 3.6 mJ, 35 fs pulses at 800 nm and 1 kHz repetition rate. 
The vertically polarized Raman pulse is synthesized by a non-collinear optical parametric amplifier (NOPA) that produces tunable pulses in the visible range (500-700 nm) and its compression is controlled by a pair of chirped mirrors~\cite{Zavelani-Rossi_2001}. The RP energy focused on the sample is  $\approx$ 600 nJ.
The 50 nJ vertically polarized probe pulse is synthesized trough supercontinuum generation by focusing part of the source pulse on a sapphire plate and filtering the 800 nm component by means of a shortpass filter. The PP chirp can be modified by introducing glass windows of different widths along the beam path and finely tuned changing the parameters of the super continuum generation inside the sapphire crystal. A convenient way to introduce a large amount of chirp in the PP is to add along the PP path a glan-laser polarizer, which ensures at the same time a parallel polarization between all the spectral components of the probe pulse~\cite{Ferrante2018}. Both the PP and the RP polarizations can be rotated using half-wave plates on the beam paths.  The time interval between Raman and probe pulses can be controlled by an optical delay line on the Raman pulse path. A synchronized chopper blocks alternating RP pulses in order to obtain the transmitted WLC probe both in presence and in absence of the Raman excitation.
Raman and probe pulses are then focused on a 500 $\mu$m thick optical glass cuvette containing the $\mathrm{CCl_{4}}$ or the  $\mathrm{CHCl_{3}}$ samples.
Finally, the PP is frequency dispersed by a spectrometer onto a CCD device.
The OKE cross-correlation was performed for measuring the temporal overlap between the different spectral components of the PP and the RP and by means of the same experimental apparatus, with an additional glan-laser polarizer, suppressing the PP transmission in absence of the Raman pulse.
For performing the frequency domain SRS measurements reported in Table \ref{table: SRS} the femtosecond RP was replaced by a narrowband pulse ($\sim$ 10 cm$^{-1}$ of bandwidth centered at 596 nm), temporally overlapped with the PP inside the sample.

\begin{acknowledgement}

G.B. acknowledges the `Avvio alla ricerca 2018' grant by Sapienza~Universit\'a~di~Roma.

\end{acknowledgement}

\begin{suppinfo}

The following files are available free of charge.
\begin{itemize}
  \item  OKE measure performed on $\mathrm{CCl_{4}}$ for calibrating the probe chirp and the temporal delay between RP and PP;
  complete derivation of the $S^{(3)}_A$ and $S^{(3)}_B$ CISRS signals; conventional ISRS spectra of $\mathrm{CCl_{4}}$ and $\mathrm{CHCl_{3}}$;
\end{itemize}

\end{suppinfo}

%\bibliography{achemso-demo}

\begin{mcitethebibliography}{47}
		\providecommand{\latin}[1]{#1}
		\makeatletter
		\providecommand{\doi}
		{\begingroup\let\do\@makeother\dospecials
			\catcode`\{=1 \catcode`\}=2 \doi@aux}
		\providecommand{\doi@aux}[1]{\endgroup\texttt{#1}}
		\makeatother
		\providecommand*\mcitethebibliography{\thebibliography}
		\csname @ifundefined\endcsname{endmcitethebibliography}
		{\let\endmcitethebibliography\endthebibliography}{}
			\providecommand*\natexlab[1]{#1}
			\providecommand*\mciteSetBstSublistMode[1]{}
			\providecommand*\mciteSetBstMaxWidthForm[2]{}
			\providecommand*\mciteBstWouldAddEndPuncttrue
			{\def\EndOfBibitem{\unskip.}}
			\providecommand*\mciteBstWouldAddEndPunctfalse
			{\let\EndOfBibitem\relax}
			\providecommand*\mciteSetBstMidEndSepPunct[3]{}
			\providecommand*\mciteSetBstSublistLabelBeginEnd[3]{}
			\providecommand*\EndOfBibitem{}
			\mciteSetBstSublistMode{f}
			\mciteSetBstMaxWidthForm{subitem}{(\alph{mcitesubitemcount})}
			\mciteSetBstSublistLabelBeginEnd
			{\mcitemaxwidthsubitemform\space}
			{\relax}
			{\relax}
			
			\bibitem[Liebel \latin{et~al.}(2015)Liebel, Schnedermann, Wende, and
			Kukura]{cit::IVS::kukura}
			Liebel,~M.; Schnedermann,~C.; Wende,~T.; Kukura,~P. Principles and Applications
			of Broadband Impulsive Vibrational Spectroscopy. \emph{J. Phys. Chem. A}
			\textbf{2015}, \emph{119}, 9506--9517\relax
			\mciteBstWouldAddEndPuncttrue
			\mciteSetBstMidEndSepPunct{\mcitedefaultmidpunct}
			{\mcitedefaultendpunct}{\mcitedefaultseppunct}\relax
			\EndOfBibitem
			\bibitem[Cheng and Xie(2004)Cheng, and Xie]{Cheng2004}
			Cheng,~J.-X.; Xie,~X.~S. Coherent Anti-Stokes Raman Scattering Microscopy:~
			Instrumentation, Theory, and Applications. \emph{J. Phys. Chem. B}
			\textbf{2004}, \emph{108}, 827--840\relax
			\mciteBstWouldAddEndPuncttrue
			\mciteSetBstMidEndSepPunct{\mcitedefaultmidpunct}
			{\mcitedefaultendpunct}{\mcitedefaultseppunct}\relax
			\EndOfBibitem
			\bibitem[Batignani \latin{et~al.}(2019)Batignani, Fumero, Pontecorvo, Ferrante,
			Mukamel, and Scopigno]{Batignani2019}
			Batignani,~G.; Fumero,~G.; Pontecorvo,~E.; Ferrante,~C.; Mukamel,~S.;
			Scopigno,~T. Genuine dynamics vs cross phase modulation artefacts in
			Femtosecond Stimulated Raman Spectroscopy. \emph{{ACS} Photonics}
			\textbf{2019}, \relax
			\mciteBstWouldAddEndPunctfalse
			\mciteSetBstMidEndSepPunct{\mcitedefaultmidpunct}
			{}{\mcitedefaultseppunct}\relax
			\EndOfBibitem
			\bibitem[Hamaguchi and Gustafson(1994)Hamaguchi, and Gustafson]{Hamaguchi1994}
			Hamaguchi,~H.; Gustafson,~T.~L. Ultrafast Time-Resolved Spontaneous and
			Coherent Raman Spectroscopy: The Structure and Dynamics of Photogenerated
			Transient Species. \emph{Annu. Rev. Phys. Chem.} \textbf{1994}, \emph{45},
			593--622\relax
			\mciteBstWouldAddEndPuncttrue
			\mciteSetBstMidEndSepPunct{\mcitedefaultmidpunct}
			{\mcitedefaultendpunct}{\mcitedefaultseppunct}\relax
			\EndOfBibitem
			\bibitem[Baskin and Zewail(2001)Baskin, and Zewail]{Zewail_nobel}
			Baskin,~J.~S.; Zewail,~A.~H. Freezing Atoms in Motion: Principles of
			Femtochemistry and Demonstration by Laser Stroboscopy. \emph{J. Chem. Educ.}
			\textbf{2001}, \relax
			\mciteBstWouldAddEndPunctfalse
			\mciteSetBstMidEndSepPunct{\mcitedefaultmidpunct}
			{}{\mcitedefaultseppunct}\relax
			\EndOfBibitem
			\bibitem[Kahan \latin{et~al.}(2007)Kahan, Nahmias, Friedman, Sheves, and
			Ruhman]{cit::rhuman::bacteriorhodopsin}
			Kahan,~A.; Nahmias,~O.; Friedman,~N.; Sheves,~M.; Ruhman,~S. Following
			Photoinduced Dynamics in Bacteriorhodopsin with 7-fs Impulsive Vibrational
			Spectroscopy. \emph{J. Am. Chem. Soc.} \textbf{2007}, \emph{129},
			537--546\relax
			\mciteBstWouldAddEndPuncttrue
			\mciteSetBstMidEndSepPunct{\mcitedefaultmidpunct}
			{\mcitedefaultendpunct}{\mcitedefaultseppunct}\relax
			\EndOfBibitem
			\bibitem[Schnedermann \latin{et~al.}(2016)Schnedermann, Muders, Ehrenberg,
			Schlesinger, Kukura, and Heberle]{Schnedermann2016}
			Schnedermann,~C.; Muders,~V.; Ehrenberg,~D.; Schlesinger,~R.; Kukura,~P.;
			Heberle,~J. Vibronic Dynamics of the Ultrafast all- trans to 13- cis
			Photoisomerization of Retinal in Channelrhodopsin-1. \emph{J. Am. Chem. Soc.}
			\textbf{2016}, \emph{138}, 4757--4762\relax
			\mciteBstWouldAddEndPuncttrue
			\mciteSetBstMidEndSepPunct{\mcitedefaultmidpunct}
			{\mcitedefaultendpunct}{\mcitedefaultseppunct}\relax
			\EndOfBibitem
			\bibitem[Ruhman \latin{et~al.}(1987)Ruhman, Kohler, Joly, and
			Nelson]{cit::rhuman::liquid}
			Ruhman,~S.; Kohler,~B.; Joly,~A.~G.; Nelson,~K.~A. Intermolecular vibrational
			motion in {CS}$_2$ liquid at 165 $\leqslant$ T$\geqslant$ 300 {K} observed by
			femtosecond time-resolved impulsive stimulated scattering. \emph{Chem. Phys.
				Let.} \textbf{1987}, \emph{141}, 16--24\relax
			\mciteBstWouldAddEndPuncttrue
			\mciteSetBstMidEndSepPunct{\mcitedefaultmidpunct}
			{\mcitedefaultendpunct}{\mcitedefaultseppunct}\relax
			\EndOfBibitem
			\bibitem[Fujisawa \latin{et~al.}(2016)Fujisawa, Kuramochi, Hosoi, Takeuchi, and
			Tahara]{Fujisawa2016}
			Fujisawa,~T.; Kuramochi,~H.; Hosoi,~H.; Takeuchi,~S.; Tahara,~T. Role of
			Coherent Low-Frequency Motion in Excited-State Proton Transfer of Green
			Fluorescent Protein Studied by Time-Resolved Impulsive Stimulated Raman
			Spectroscopy. \emph{J. Am. Chem. Soc.} \textbf{2016}, \emph{138},
			3942--3945\relax
			\mciteBstWouldAddEndPuncttrue
			\mciteSetBstMidEndSepPunct{\mcitedefaultmidpunct}
			{\mcitedefaultendpunct}{\mcitedefaultseppunct}\relax
			\EndOfBibitem
			\bibitem[Liebel and Kukura(2013)Liebel, and Kukura]{Liebel2013}
			Liebel,~M.; Kukura,~P. Broad-Band Impulsive Vibrational Spectroscopy of Excited
			Electronic States in the Time Domain. \emph{J. Phys. Chem. Lett.}
			\textbf{2013}, \emph{4}, 1358--1364\relax
			\mciteBstWouldAddEndPuncttrue
			\mciteSetBstMidEndSepPunct{\mcitedefaultmidpunct}
			{\mcitedefaultendpunct}{\mcitedefaultseppunct}\relax
			\EndOfBibitem
			\bibitem[Kuramochi \latin{et~al.}(2017)Kuramochi, Takeuchi, Yonezawa, Kamikubo,
			Kataoka, and Tahara]{Kuramochi2017}
			Kuramochi,~H.; Takeuchi,~S.; Yonezawa,~K.; Kamikubo,~H.; Kataoka,~M.;
			Tahara,~T. Probing the early stages of photoreception in photoactive yellow
			protein with ultrafast time-domain Raman spectroscopy. \emph{Nat. Chem.}
			\textbf{2017}, \emph{9}, 660--666\relax
			\mciteBstWouldAddEndPuncttrue
			\mciteSetBstMidEndSepPunct{\mcitedefaultmidpunct}
			{\mcitedefaultendpunct}{\mcitedefaultseppunct}\relax
			\EndOfBibitem
			\bibitem[Maiuri \latin{et~al.}(2018)Maiuri, Ostroumov, Saer, Blankenship, and
			Scholes]{Maiuri_2018}
			Maiuri,~M.; Ostroumov,~E.~E.; Saer,~R.~G.; Blankenship,~R.~E.; Scholes,~G.~D.
			Coherent wavepackets in the Fenna{\textendash}Matthews{\textendash}Olson
			complex are robust to excitonic-structure perturbations caused by
			mutagenesis. \emph{Nat. Chem.} \textbf{2018}, \emph{10}, 177--183\relax
			\mciteBstWouldAddEndPuncttrue
			\mciteSetBstMidEndSepPunct{\mcitedefaultmidpunct}
			{\mcitedefaultendpunct}{\mcitedefaultseppunct}\relax
			\EndOfBibitem
			\bibitem[Musser \latin{et~al.}(2015)Musser, Liebel, Schnedermann, Wende, Kehoe,
			Rao, and Kukura]{Musser2015}
			Musser,~A.~J.; Liebel,~M.; Schnedermann,~C.; Wende,~T.; Kehoe,~T.~B.; Rao,~A.;
			Kukura,~P. Evidence for conical intersection dynamics mediating ultrafast
			singlet exciton fission. \emph{Nat. Phys.} \textbf{2015}, \emph{11},
			352--357\relax
			\mciteBstWouldAddEndPuncttrue
			\mciteSetBstMidEndSepPunct{\mcitedefaultmidpunct}
			{\mcitedefaultendpunct}{\mcitedefaultseppunct}\relax
			\EndOfBibitem
			\bibitem[Ghosh \latin{et~al.}(2017)Ghosh, Aharon, Etgar, and Ruhman]{Ghosh2017}
			Ghosh,~T.; Aharon,~S.; Etgar,~L.; Ruhman,~S. Free Carrier Emergence and Onset
			of Electron{\textendash}Phonon Coupling in Methylammonium Lead Halide
			Perovskite Films. \emph{J. Am. Chem. Soc.} \textbf{2017}, \emph{139},
			18262--18270\relax
			\mciteBstWouldAddEndPuncttrue
			\mciteSetBstMidEndSepPunct{\mcitedefaultmidpunct}
			{\mcitedefaultendpunct}{\mcitedefaultseppunct}\relax
			\EndOfBibitem
			\bibitem[Batignani \latin{et~al.}(2018)Batignani, Fumero, Kandada, Cerullo,
			Gandini, Ferrante, Petrozza, and Scopigno]{Batignani2018}
			Batignani,~G.; Fumero,~G.; Kandada,~A. R.~S.; Cerullo,~G.; Gandini,~M.;
			Ferrante,~C.; Petrozza,~A.; Scopigno,~T. Probing femtosecond lattice
			displacement upon photo-carrier generation in lead halide perovskite.
			\emph{Nat. Commun.} \textbf{2018}, \emph{9}\relax
			\mciteBstWouldAddEndPuncttrue
			\mciteSetBstMidEndSepPunct{\mcitedefaultmidpunct}
			{\mcitedefaultendpunct}{\mcitedefaultseppunct}\relax
			\EndOfBibitem
			\bibitem[Park \latin{et~al.}(2018)Park, Neukirch, Reyes-Lillo, Lai, Ellis,
			Dietze, Neaton, Yang, Tretiak, and Mathies]{Park2018}
			Park,~M.; Neukirch,~A.~J.; Reyes-Lillo,~S.~E.; Lai,~M.; Ellis,~S.~R.;
			Dietze,~D.; Neaton,~J.~B.; Yang,~P.; Tretiak,~S.; Mathies,~R.~A.
			Excited-state vibrational dynamics toward the polaron in methylammonium lead
			iodide perovskite. \emph{Nat. Commun.} \textbf{2018}, \emph{9}\relax
			\mciteBstWouldAddEndPuncttrue
			\mciteSetBstMidEndSepPunct{\mcitedefaultmidpunct}
			{\mcitedefaultendpunct}{\mcitedefaultseppunct}\relax
			\EndOfBibitem
			\bibitem[Tanimura and Mukamel(1993)Tanimura, and Mukamel]{Tanimura1993}
			Tanimura,~Y.; Mukamel,~S. Two-dimensional femtosecond vibrational spectroscopy
			of liquids. \emph{J. Chem. Phys} \textbf{1993}, \emph{99}, 9496--9511\relax
			\mciteBstWouldAddEndPuncttrue
			\mciteSetBstMidEndSepPunct{\mcitedefaultmidpunct}
			{\mcitedefaultendpunct}{\mcitedefaultseppunct}\relax
			\EndOfBibitem
			\bibitem[Tokmakoff \latin{et~al.}(1997)Tokmakoff, Lang, Larsen, Fleming,
			Chernyak, and Mukamel]{Tokmakoff1997}
			Tokmakoff,~A.; Lang,~M.~J.; Larsen,~D.~S.; Fleming,~G.~R.; Chernyak,~V.;
			Mukamel,~S. Two-Dimensional Raman Spectroscopy of Vibrational Interactions in
			Liquids. \emph{Phys. Rev. Lett.} \textbf{1997}, \emph{79}, 2702--2705\relax
			\mciteBstWouldAddEndPuncttrue
			\mciteSetBstMidEndSepPunct{\mcitedefaultmidpunct}
			{\mcitedefaultendpunct}{\mcitedefaultseppunct}\relax
			\EndOfBibitem
			\bibitem[Molesky \latin{et~al.}(2016)Molesky, Guo, Cheshire, and
			Moran]{Moran2016_perspective}
			Molesky,~B.~P.; Guo,~Z.; Cheshire,~T.~P.; Moran,~A.~M. Perspective:
			Two-dimensional resonance Raman spectroscopy. \emph{J. Chem. Phys}
			\textbf{2016}, \emph{145}, 180901\relax
			\mciteBstWouldAddEndPuncttrue
			\mciteSetBstMidEndSepPunct{\mcitedefaultmidpunct}
			{\mcitedefaultendpunct}{\mcitedefaultseppunct}\relax
			\EndOfBibitem
			\bibitem[Kuramochi \latin{et~al.}(2019)Kuramochi, Takeuchi, Kamikubo, Kataoka,
			and Tahara]{Kuramochieaau4490}
			Kuramochi,~H.; Takeuchi,~S.; Kamikubo,~H.; Kataoka,~M.; Tahara,~T. Fifth-order
			time-domain Raman spectroscopy of photoactive yellow protein for visualizing
			vibrational coupling in its excited state. \emph{Sci. Adv.} \textbf{2019},
			\emph{5}\relax
			\mciteBstWouldAddEndPuncttrue
			\mciteSetBstMidEndSepPunct{\mcitedefaultmidpunct}
			{\mcitedefaultendpunct}{\mcitedefaultseppunct}\relax
			\EndOfBibitem
			\bibitem[Hellerer \latin{et~al.}(2004)Hellerer, Enejder, and
			Zumbusch]{Hellerer2004}
			Hellerer,~T.; Enejder,~A.~M.; Zumbusch,~A. Spectral focusing: High spectral
			resolution spectroscopy with broad-bandwidth laser pulses. \emph{Appl. Phys.
				Lett.} \textbf{2004}, \emph{85}, 25--27\relax
			\mciteBstWouldAddEndPuncttrue
			\mciteSetBstMidEndSepPunct{\mcitedefaultmidpunct}
			{\mcitedefaultendpunct}{\mcitedefaultseppunct}\relax
			\EndOfBibitem
			\bibitem[Rocha-Mendoza \latin{et~al.}(2008)Rocha-Mendoza, Langbein, and
			Borri]{RochaMendoza2008}
			Rocha-Mendoza,~I.; Langbein,~W.; Borri,~P. Coherent anti-Stokes Raman
			microspectroscopy using spectral focusing with glass dispersion. \emph{Appl.
				Phys. Lett.} \textbf{2008}, \emph{93}, 201103\relax
			\mciteBstWouldAddEndPuncttrue
			\mciteSetBstMidEndSepPunct{\mcitedefaultmidpunct}
			{\mcitedefaultendpunct}{\mcitedefaultseppunct}\relax
			\EndOfBibitem
			\bibitem[Rocha-Mendoza \latin{et~al.}(2009)Rocha-Mendoza, Langbein, Watson, and
			Borri]{RochaMendoza2009}
			Rocha-Mendoza,~I.; Langbein,~W.; Watson,~P.; Borri,~P. Differential coherent
			anti-Stokes Raman scattering microscopy with linearly chirped femtosecond
			laser pulses. \emph{Opt. Lett.} \textbf{2009}, \emph{34}, 2258\relax
			\mciteBstWouldAddEndPuncttrue
			\mciteSetBstMidEndSepPunct{\mcitedefaultmidpunct}
			{\mcitedefaultendpunct}{\mcitedefaultseppunct}\relax
			\EndOfBibitem
			\bibitem[Saltarelli \latin{et~al.}(2018)Saltarelli, Kumar, Viola, Crisafi,
			Perri, Cerullo, and Polli]{Saltarelli2018}
			Saltarelli,~F.; Kumar,~V.; Viola,~D.; Crisafi,~F.; Perri,~A.; Cerullo,~G.;
			Polli,~D. Photonic time stretch to measure small spectral changes with broad
			wavelength coverage for high-speed coherent Raman microscopy (Conference
			Presentation). Real-time Measurements, Rogue Phenomena, and Single-Shot
			Applications {III}. 2018\relax
			\mciteBstWouldAddEndPuncttrue
			\mciteSetBstMidEndSepPunct{\mcitedefaultmidpunct}
			{\mcitedefaultendpunct}{\mcitedefaultseppunct}\relax
			\EndOfBibitem
			\bibitem[Bardeen \latin{et~al.}(1995)Bardeen, Wang, and Shank]{Bardeen1995}
			Bardeen,~C.~J.; Wang,~Q.; Shank,~C.~V. Selective Excitation of Vibrational Wave
			Packet Motion Using Chirped Pulses. \emph{Phys. Rev. Lett.} \textbf{1995},
			\emph{75}, 3410--3413\relax
			\mciteBstWouldAddEndPuncttrue
			\mciteSetBstMidEndSepPunct{\mcitedefaultmidpunct}
			{\mcitedefaultendpunct}{\mcitedefaultseppunct}\relax
			\EndOfBibitem
			\bibitem[Wand \latin{et~al.}(2010)Wand, Kallush, Shoshanim, Bismuth, Kosloff,
			and Ruhman]{cit::RuhmanWand}
			Wand,~A.; Kallush,~S.; Shoshanim,~O.; Bismuth,~O.; Kosloff,~R.; Ruhman,~S.
			Chirp effects on impulsive vibrational spectroscopy: a multimode perspective.
			\emph{Phys. Chem. Chem. Phys.} \textbf{2010}, \emph{12}, 2149--2163\relax
			\mciteBstWouldAddEndPuncttrue
			\mciteSetBstMidEndSepPunct{\mcitedefaultmidpunct}
			{\mcitedefaultendpunct}{\mcitedefaultseppunct}\relax
			\EndOfBibitem
			\bibitem[Monacelli \latin{et~al.}(2017)Monacelli, Batignani, Fumero, Ferrante,
			Mukamel, and Scopigno]{Monacelli2017}
			Monacelli,~L.; Batignani,~G.; Fumero,~G.; Ferrante,~C.; Mukamel,~S.;
			Scopigno,~T. Manipulating Impulsive Stimulated Raman Spectroscopy with a
			Chirped Probe Pulse. \emph{J. Phys. Chem. Lett.} \textbf{2017}, \emph{8},
			966--974\relax
			\mciteBstWouldAddEndPuncttrue
			\mciteSetBstMidEndSepPunct{\mcitedefaultmidpunct}
			{\mcitedefaultendpunct}{\mcitedefaultseppunct}\relax
			\EndOfBibitem
			\bibitem[Gdor \latin{et~al.}(2017)Gdor, Ghosh, Lioubashevski, and
			Ruhman]{Gdor2017}
			Gdor,~I.; Ghosh,~T.; Lioubashevski,~O.; Ruhman,~S. Nonresonant Raman Effects on
			Femtosecond Pump{\textendash}Probe with Chirped White Light: Challenges and
			Opportunities. \emph{J. Phys. Chem. Lett.} \textbf{2017}, \emph{8},
			1920--1924\relax
			\mciteBstWouldAddEndPuncttrue
			\mciteSetBstMidEndSepPunct{\mcitedefaultmidpunct}
			{\mcitedefaultendpunct}{\mcitedefaultseppunct}\relax
			\EndOfBibitem
			\bibitem[Strickland and Mourou(1985)Strickland, and Mourou]{Strickland1985}
			Strickland,~D.; Mourou,~G. Compression of amplified chirped optical pulses.
			\emph{Opt. Commun.} \textbf{1985}, \emph{56}, 219--221\relax
			\mciteBstWouldAddEndPuncttrue
			\mciteSetBstMidEndSepPunct{\mcitedefaultmidpunct}
			{\mcitedefaultendpunct}{\mcitedefaultseppunct}\relax
			\EndOfBibitem
			\bibitem[Teo \latin{et~al.}(2015)Teo, Ofori-Okai, Werley, and Nelson]{Teo2015}
			Teo,~S.~M.; Ofori-Okai,~B.~K.; Werley,~C.~A.; Nelson,~K.~A. Invited Article:
			Single-shot {THz} detection techniques optimized for multidimensional {THz}
			spectroscopy. \emph{Rev. Sci. Instrum.} \textbf{2015}, \emph{86},
			051301\relax
			\mciteBstWouldAddEndPuncttrue
			\mciteSetBstMidEndSepPunct{\mcitedefaultmidpunct}
			{\mcitedefaultendpunct}{\mcitedefaultseppunct}\relax
			\EndOfBibitem
			\bibitem[Waldecker \latin{et~al.}(2015)Waldecker, Miller, Rud{\'{e}}, Bertoni,
			Osmond, Pruneri, Simpson, Ernstorfer, and Wall]{Waldecker2015}
			Waldecker,~L.; Miller,~T.~A.; Rud{\'{e}},~M.; Bertoni,~R.; Osmond,~J.;
			Pruneri,~V.; Simpson,~R.~E.; Ernstorfer,~R.; Wall,~S. Time-domain separation
			of optical properties from structural transitions in resonantly
			bonded~materials. \emph{Nat. Mater.} \textbf{2015}, \emph{14}, 991--995\relax
			\mciteBstWouldAddEndPuncttrue
			\mciteSetBstMidEndSepPunct{\mcitedefaultmidpunct}
			{\mcitedefaultendpunct}{\mcitedefaultseppunct}\relax
			\EndOfBibitem
			\bibitem[Polli \latin{et~al.}(2010)Polli, Brida, Mukamel, Lanzani, and
			Cerullo]{Polli2010}
			Polli,~D.; Brida,~D.; Mukamel,~S.; Lanzani,~G.; Cerullo,~G. Effective temporal
			resolution in pump-probe spectroscopy with strongly chirped pulses.
			\emph{Phys. Rev. A} \textbf{2010}, \emph{82}, 053809\relax
			\mciteBstWouldAddEndPuncttrue
			\mciteSetBstMidEndSepPunct{\mcitedefaultmidpunct}
			{\mcitedefaultendpunct}{\mcitedefaultseppunct}\relax
			\EndOfBibitem
			\bibitem[Agrawal(2013)]{cit::Agrawal}
			Agrawal,~G. \emph{Nonlinear Fiber Optics}; Academic Press, 2013\relax
			\mciteBstWouldAddEndPuncttrue
			\mciteSetBstMidEndSepPunct{\mcitedefaultmidpunct}
			{\mcitedefaultendpunct}{\mcitedefaultseppunct}\relax
			\EndOfBibitem
			\bibitem[Mukamel(1995)]{cit::Mukamel}
			Mukamel,~S. \emph{Principles of nonlinear spectroscopy}; Oxford university
			press, 1995\relax
			\mciteBstWouldAddEndPuncttrue
			\mciteSetBstMidEndSepPunct{\mcitedefaultmidpunct}
			{\mcitedefaultendpunct}{\mcitedefaultseppunct}\relax
			\EndOfBibitem
			\bibitem[Rahav and Mukamel(2010)Rahav, and Mukamel]{Mukamel_Rahav}
			Rahav,~S.; Mukamel,~S. Ultrafast Nonlinear Optical Signals Viewed from the
			Molecule's Perspective: Kramers-{H}eisenberg Transition-Amplitudes versus
			Susceptibilities. \emph{Adv. At., Mol., Opt. Phys.} \textbf{2010}, \emph{59},
			223--263\relax
			\mciteBstWouldAddEndPuncttrue
			\mciteSetBstMidEndSepPunct{\mcitedefaultmidpunct}
			{\mcitedefaultendpunct}{\mcitedefaultseppunct}\relax
			\EndOfBibitem
			\bibitem[Dorfman \latin{et~al.}(2013)Dorfman, Fingerhut, and
			Mukamel]{Dorfman2013}
			Dorfman,~K.~E.; Fingerhut,~B.~P.; Mukamel,~S. Time-resolved broadband Raman
			spectroscopies: A unified six-wave-mixing representation. \emph{J. Chem.
				Phys} \textbf{2013}, \emph{139}, 124113\relax
			\mciteBstWouldAddEndPuncttrue
			\mciteSetBstMidEndSepPunct{\mcitedefaultmidpunct}
			{\mcitedefaultendpunct}{\mcitedefaultseppunct}\relax
			\EndOfBibitem
			\bibitem[Batignani \latin{et~al.}(2015)Batignani, Fumero, Mukamel, and
			Scopigno]{Batignani2015pccp}
			Batignani,~G.; Fumero,~G.; Mukamel,~S.; Scopigno,~T. Energy flow between
			spectral components in 2D broadband stimulated Raman spectroscopy.
			\emph{Phys. Chem. Chem. Phys.} \textbf{2015}, \emph{17}, 10454--10461\relax
			\mciteBstWouldAddEndPuncttrue
			\mciteSetBstMidEndSepPunct{\mcitedefaultmidpunct}
			{\mcitedefaultendpunct}{\mcitedefaultseppunct}\relax
			\EndOfBibitem
			\bibitem[Batignani \latin{et~al.}(2016)Batignani, Pontecorvo, Ferrante, Aschi,
			Elles, and Scopigno]{Batignani2016}
			Batignani,~G.; Pontecorvo,~E.; Ferrante,~C.; Aschi,~M.; Elles,~C.~G.;
			Scopigno,~T. Visualizing Excited-State Dynamics of a Diaryl Thiophene:
			Femtosecond Stimulated Raman Scattering as a Probe of Conjugated Molecules.
			\emph{J. Phys. Chem. Lett.} \textbf{2016}, \emph{7}, 2981--2988\relax
			\mciteBstWouldAddEndPuncttrue
			\mciteSetBstMidEndSepPunct{\mcitedefaultmidpunct}
			{\mcitedefaultendpunct}{\mcitedefaultseppunct}\relax
			\EndOfBibitem
			\bibitem[Batignani \latin{et~al.}(2016)Batignani, Pontecorvo, Giovannetti,
			Ferrante, Fumero, and Scopigno]{Batignani2016SciRep}
			Batignani,~G.; Pontecorvo,~E.; Giovannetti,~G.; Ferrante,~C.; Fumero,~G.;
			Scopigno,~T. Electronic resonances in broadband stimulated Raman
			spectroscopy. \emph{Sci. Rep.} \textbf{2016}, \emph{6}\relax
			\mciteBstWouldAddEndPuncttrue
			\mciteSetBstMidEndSepPunct{\mcitedefaultmidpunct}
			{\mcitedefaultendpunct}{\mcitedefaultseppunct}\relax
			\EndOfBibitem
			\bibitem[McCamant \latin{et~al.}(2004)McCamant, Kukura, Yoon, and
			Mathies]{McCamant2004}
			McCamant,~D.~W.; Kukura,~P.; Yoon,~S.; Mathies,~R.~A. Femtosecond broadband
			stimulated Raman spectroscopy: Apparatus and methods. \emph{Rev. Sci.
				Instrum.} \textbf{2004}, \emph{75}, 4971--4980\relax
			\mciteBstWouldAddEndPuncttrue
			\mciteSetBstMidEndSepPunct{\mcitedefaultmidpunct}
			{\mcitedefaultendpunct}{\mcitedefaultseppunct}\relax
			\EndOfBibitem
			\bibitem[Shimanouchi(1972)]{book_Shimanouchi}
			Shimanouchi,~T. \emph{Tables of molecular vibrational frequencies. Consolidated
				volume I}; National Bureau of Standards, 1972\relax
			\mciteBstWouldAddEndPuncttrue
			\mciteSetBstMidEndSepPunct{\mcitedefaultmidpunct}
			{\mcitedefaultendpunct}{\mcitedefaultseppunct}\relax
			\EndOfBibitem
			\bibitem[Marrinan and Sheppard(1954)Marrinan, and Sheppard]{Marrinan1954}
			Marrinan,~H.~J.; Sheppard,~N. Relative Intensities of the Raman Lines of Carbon
			Tetrachloride, Chloroform, and Methylene Chloride. \emph{J. Opt. Soc. Am.}
			\textbf{1954}, \emph{44}, 815\relax
			\mciteBstWouldAddEndPuncttrue
			\mciteSetBstMidEndSepPunct{\mcitedefaultmidpunct}
			{\mcitedefaultendpunct}{\mcitedefaultseppunct}\relax
			\EndOfBibitem
			\bibitem[Gaynor \latin{et~al.}(2015)Gaynor, Wetterer, Cochran, Valente, and
			Mayer]{Gaynor2015}
			Gaynor,~J.~D.; Wetterer,~A.~M.; Cochran,~R.~M.; Valente,~E.~J.; Mayer,~S.~G.
			Vibrational Spectroscopy of the {CCl}4 $\nu$1 Mode: Effect of Thermally
			Populated Vibrational States. \emph{J. Chem. Educ.} \textbf{2015}, \emph{92},
			1949--1952\relax
			\mciteBstWouldAddEndPuncttrue
			\mciteSetBstMidEndSepPunct{\mcitedefaultmidpunct}
			{\mcitedefaultendpunct}{\mcitedefaultseppunct}\relax
			\EndOfBibitem
			\bibitem[McCreery(2000)]{McCreery2000}
			McCreery,~R.~L. \emph{Raman Spectroscopy for Chemical Analysis}; John Wiley
			{\&} Sons, Inc., 2000\relax
			\mciteBstWouldAddEndPuncttrue
			\mciteSetBstMidEndSepPunct{\mcitedefaultmidpunct}
			{\mcitedefaultendpunct}{\mcitedefaultseppunct}\relax
			\EndOfBibitem
			\bibitem[Zavelani-Rossi \latin{et~al.}(2001)Zavelani-Rossi, Cerullo, Silvestri,
			Gallmann, Matuschek, Steinmeyer, Keller, Angelow, Scheuer, and
			Tschudi]{Zavelani-Rossi_2001}
			Zavelani-Rossi,~M.; Cerullo,~G.; Silvestri,~S.~D.; Gallmann,~L.; Matuschek,~N.;
			Steinmeyer,~G.; Keller,~U.; Angelow,~G.; Scheuer,~V.; Tschudi,~T. Pulse
			compression over a 170-THz bandwidth in the visible by use of only chirped
			mirrors. \emph{Opt. Lett.} \textbf{2001}, \emph{26}, 1155--1157\relax
			\mciteBstWouldAddEndPuncttrue
			\mciteSetBstMidEndSepPunct{\mcitedefaultmidpunct}
			{\mcitedefaultendpunct}{\mcitedefaultseppunct}\relax
			\EndOfBibitem
			\bibitem[Ferrante \latin{et~al.}(2018)Ferrante, Batignani, Fumero, Pontecorvo,
			Virga, Montemiglio, Cerullo, Vos, and Scopigno]{Ferrante2018}
			Ferrante,~C.; Batignani,~G.; Fumero,~G.; Pontecorvo,~E.; Virga,~A.;
			Montemiglio,~L.~C.; Cerullo,~G.; Vos,~M.~H.; Scopigno,~T. Resonant broadband
			stimulated Raman scattering in myoglobin. \emph{J. Raman Spectrosc.}
			\textbf{2018}, \relax
			\mciteBstWouldAddEndPunctfalse
			\mciteSetBstMidEndSepPunct{\mcitedefaultmidpunct}
			{}{\mcitedefaultseppunct}\relax
			\EndOfBibitem
		\end{mcitethebibliography}

\end{document}